\documentclass{emulateapj}

\slugcomment{\today, accepted for publication in ApJL}

\shorttitle{The Radio Jet of PKS~0637--752}
\shortauthors{Godfrey et al.}

\begin{document}

\title{Periodic structure in the Mpc-scale jet of PKS~0637--752}

\author{L.E.H. Godfrey\altaffilmark{1,2},  J.E.J. Lovell\altaffilmark{3,4}, S. Burke-Spolaor\altaffilmark{5}, R. Ekers\altaffilmark{3,1}, G.V. Bicknell\altaffilmark{2}, M.~Birkinshaw\altaffilmark{6,7}, D. M. Worrall\altaffilmark{6,7}, D.L. Jauncey\altaffilmark{3,2},   D. A. Schwartz\altaffilmark{6}, H. L. Marshall\altaffilmark{8},  J. Gelbord\altaffilmark{9},  E.S.~Perlman\altaffilmark{10},
   M. Georganopoulos\altaffilmark{11}}

\email{L.Godfrey@curtin.edu.au}

\altaffiltext{1}{International Centre for Radio Astronomy Research, Curtin University, GPO Box U1987, Perth, WA, 6102, Australia}
\altaffiltext{2}{Research School of Astronomy and Astrophysics, Australian National
University, Cotter Road, Weston, ACT, 2611, Australia}
\altaffiltext{3}{CSIRO Astronomy and Space Science, Australia Telescope National Facility, Epping, NSW1710, Australia}
%\altaffiltext{4}{CSIRO, Industrial Physics, PO Box 218 Lindfield NSW 2070, Australia}
\altaffiltext{4}{School of Mathematics and Physics, University of Tasmania, Tas 7001, Australia}
\altaffiltext{5}{NASA Jet Propulsion Laboratory, 4800 Oak Grove Drive, Pasadena, CA 91109, USA}
\altaffiltext{6}{Harvard-Smithsonian Center for Astrophysics, 60 Garden Street, Cambridge, MA 02138, USA}
\altaffiltext{7}{HH Wills Physics Laboratory, University of Bristol, Tyndall Avenue, Bristol BS8 1TL, UK}
\altaffiltext{8}{Kavli Institute for Astrophysics and Space Research, Massachusetts Institute of Technology, USA}
\altaffiltext{9}{Department of Physics, Durham University, South Road, Durham, DH1 3LE, UK}
\altaffiltext{10}{Physics and Space Sciences Department, Florida Institute of Technology, 150 West University Boulevard, Melbourne, FL 32901, USA}
\altaffiltext{11}{Department of Physics, Joint Center for Astrophysics, University of Maryland-Baltimore County, 1000 Hilltop Circle, Baltimore, MD 21250, USA}


\begin{abstract}

We present 18~GHz Australia Telescope Compact Array imaging of the Mpc-scale quasar jet PKS~0637--752 with angular resolution $\sim 0\farcs58$. We draw attention to a spectacular train of quasi-periodic knots along the inner 11$\arcsec$ of the jet, with average separation $\sim$ 1.1 arcsec (7.6 kpc projected). We consider two classes of model to explain the periodic knots: those that involve a static pattern through which the jet plasma travels (e.\,g. stationary shocks); and those that involve modulation of the jet engine.  Interpreting the knots as re-confinement shocks implies the jet kinetic power $Q_{\rm jet} \sim10^{46}$~ergs~s$^{-1}$, but the constant knot separation along the jet is not expected in a realistic external density profile. For models involving modulation of the jet engine, we find that the required modulation period is $2 \times 10^3 ~ {\rm yr} < \tau < 3 \times 10^5 ~ {\rm yr}$. The lower end of this range is applicable if the jet remains highly relativistic on kpc-scales, as implied by the IC/CMB model of jet X-ray emission. 
We suggest that the periodic jet structure in PKS~0637--752 may be analogous to the quasi-periodic jet modulation seen in the microquasar GRS~1915+105, believed to result from limit cycle behaviour in an unstable accretion disk. If variations in the accretion rate are driven by a binary black hole, the predicted orbital radius is $0.7\lesssim a \lesssim 30$~pc, which corresponds to a maximum angular separation of $\sim$0.1--5~mas.

%\textbf{This may be a scaled up version of GRS1915+105, which shows quasi-periodic variability in its radio light curve on a time-scale of $\sim 30$ minutes. Scaling this timescale by the ratio of black hole masses ($\sim \frac{10^9}{10}$) we obtain something in the order of 10$^4$ years.}
 
\end{abstract}

\section{Introduction}

\begin{figure*}[ht!]
\centering
\includegraphics[scale=0.5, angle=270]{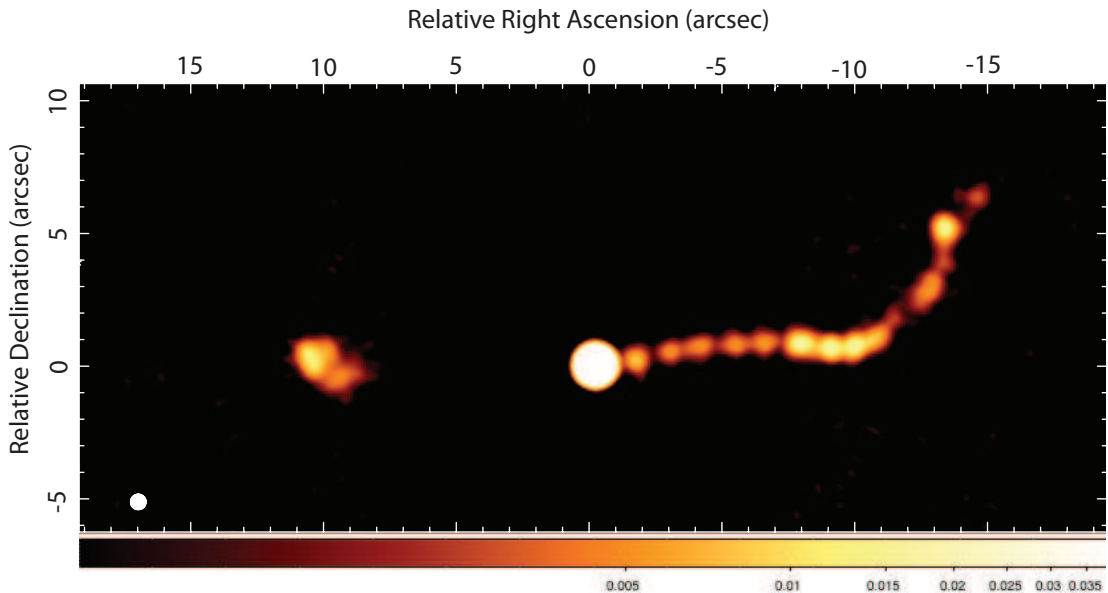}
\caption{: ATCA image of PKS~0637--752 at 17.7 GHz. The colour scale is logarithmic between a minimum and maximum surface brightness of 0.5 and 38 mJy/beam respectively. This colour scale was chosen to show clearly the regular nature of the inner jet knots. The restoring beam is circular with FWHM = $0\farcs575$. The dynamic range is 16,000:1; the off-source rms noise level is 0.3 mJy/beam, and the peak of the faintest knot in the inner jet is 5 mJy/beam. Knots in the inner jet are real and are well above the side-lobes from the bright core which are at or below the noise level.
\label{fig:17GHz_colourmap}}
\end{figure*}

The jets of FRII radio galaxies and quasars often exhibit regions of enhanced brightness, commonly referred to as ``knots". The production of knots and the associated particle acceleration mechanism are poorly understood \citep[e.\,g.][]{stawarz04a, niemiec06}. Jet knots have traditionally been identified with strong shocks in a uniform and continuous flow. In this class of models, particles are accelerated at the shock via the first order Fermi mechanism, resulting in a power-law electron energy distribution. Such internal shocks could be formed due to intrinsic velocity irregularities in the jet \citep[][]{rees78}, they could be re-confinement/re-collimation shocks \citep[][]{komissarov98}, or they could be associated with large scale instabilities in the flow \citep[e.\,g.][]{bicknell96}. Alternatively, knots in large-scale, powerful radio sources may represent moving and separate portions of the jet matter with higher speed and/or luminosity \citep{stawarz04a}. 

The radio morphology of PKS\,0637--752 ($z=0.653$) consists of a bright core; a knotty, one-sided
radio jet extending 15 arcseconds west of the core; and a
counter-lobe 11 arcseconds east of the core (see Figure \ref{fig:17GHz_colourmap}).
The apparent superluminal motion of pc-scale jet components with $v/c \approx 13$ implies a jet viewing angle less than $9^\circ$ \citep{lovell00, edwards06}, and hence source size greater than 1 Mpc. The object has received considerable attention in the past decade, largely due to the bright X-ray emission associated with the 100-kpc-scale western jet. The strong jet X-ray emission is hard to explain in terms of standard emission mechanisms such as thermal Bremsstrahlung or synchrotron self Compton \citep{schwartz00}. As a potential solution to this problem, \citet{tavecchio00} and \citet{celotti01} proposed the now-popular beamed, equipartition inverse Compton model, in which the cosmic microwave background (CMB) provides the soft photons that are scattered to X-ray energies (IC/CMB model). 

Since the launch of \emph{Chandra} in 1999, several tens of quasar X-ray jets similar to PKS~0637--752 have been detected. Despite a decade of intense observational and theoretical analysis, the X-ray emission mechanism, dynamics, and the physical processes shaping jet morphology remain unresolved. A number of problems with the IC/CMB model have been identified \citep[see e.\,g.][]{harris06, worrall09}, and whilst none of the problems is seen as directly refuting the model, mounting criticism has meant that competing models such as dual population synchrotron models \citep[e.\,g.][]{tavecchio03, jester06} are considered as likely candidates. Additional constraints are required to determine the X-ray emission mechanism and the nature of large-scale quasar jets. In the present work, we consider a new constraint on the jet dynamics: the existence of quasi-periodic peaks in the jet emission. 
In \S \ref{sec:observations} we describe the radio observations and data reduction. In \S \ref{sec:periodic_knots} we discuss the periodic structure in the jet and possible interpretations. 

We adopt the following cosmology: $H_0 = 71$~km/s/Mpc, $\Omega_m = 0.27$, $\Omega_\Lambda = 0.73$.

\section{Observations} \label{sec:observations}

PKS~0637--752 was observed with the ATCA in the 6C configuration at 17.7~GHz and 20.2~GHz on 10 May 2004. A full 12 hour synthesis was obtained, recording 128 MHz bandwidth. Regular scans on the nearby phase calibrator 0454-810 were scheduled throughout the observations, as well as scans on the ATCA flux calibrator PKS~1934-638. Standard calibration and editing procedures were carried out using the MIRIAD data analysis package. Following the initial calibration, data were exported to DIFMAP and several imaging/self-calibration iterations were performed, incorporating both phase and amplitude self-calibration.

\section{Periodic Jet Knots} \label{sec:periodic_knots}

\begin{figure}[ht!]
\centering
\plotone{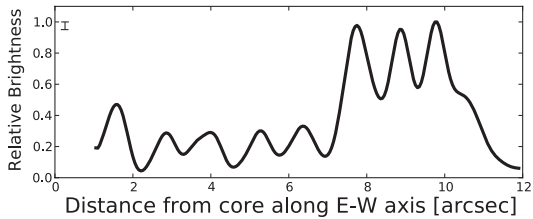}
\caption{: Plot of jet brightness integrated along the North-South direction, as a function of distance from the core in a westerly direction. The bar in the top left of the plot indicates the off-source rms of the projection. \label{fig:17GHz_profile}}
\end{figure}

Trains of bright knots, in some cases with a regular or quasi-periodic pattern, occur in radio galaxy and quasar jets: 3C 175, 3C 204, 3C 263, 3C 334 and 3C303 are prime examples \citep[][]{bridle94, kronberg86}. Figures \ref{fig:17GHz_colourmap} and \ref{fig:17GHz_profile} illustrate the periodic nature of the jet knots in PKS~0637--752, extending along the inner jet up to 11 arcseconds from the core. The average apparent knot separation is $1\farcs1$ with standard deviation $0\farcs15$, corresponding to projected separation $D_{\rm app} = 7.6 \pm 1.0$\,kpc. Following \citet{mehta09}, we argue that the jet angle to the line of sight must be greater than $\sim 4^\circ$, otherwise the deprojected source size would be uncomfortably large ($> 2$ Mpc). Based on the \citet{lovell00} component identifications and inferred superluminal motion in the pc-scale jet, \citet{edwards06} obtain a limit of $\theta < 9^\circ$ for the jet viewing angle. The parsec-scale and arcsecond-scale radio jets are well aligned \citep{lovell00, schwartz00} so that, unless the jet goes through a large bend in a plane perpendicular to the plane of sky, the jet viewing angle on arcsecond scales is likewise $< 9^{\circ}$. Thus, if the knots are stationary, the de-projected knot separation is 
\begin{equation}
50~\mbox{kpc} < D < 110~\mbox{kpc}
\end{equation}
As discussed in \S \ref{sec:variation_in_the_jet_engine}, if the knots are discrete jet components with relativistic velocity then light travel time effects must be taken into account when calculating the knot separation. 

The periodic nature of the knots in PKS~0637--752 is striking, and here we ask the question: what physical process is responsible for the periodic knot structure? We can consider the knots to be either: (A) a static pattern through which the flow travels (e.\,g. stationary shocks); or (B) a real variation within the flow caused by a variable jet engine. Section \ref{sec:stationary_pattern} deals with models that fall under category A, while Section \ref{sec:variation_in_the_jet_engine} deals with models that fall under category B. 

%We defer more detailed investigation involving spectral index and polarisation maps to an upcoming publication, based on more recent ATCA imaging. 

We defer a more detailed investigation to an upcoming publication using spectral index and polarisation data based on more sensitive, and higher resolution, ongoing ATCA imaging.

\subsection{Large-scale shocks in a continuous flow}  \label{sec:stationary_pattern} 

\subsubsection{Re-confinement Shocks}

Hydrostatic jet confinement is accompanied by the formation of regularly spaced shocks along the jet \citep{sanders83, komissarov98}, which could explain the periodic radio structure of PKS~0637--752. Regularly spaced trains of re-confinement shocks are routinely observed in hydrodynamic simulations of AGN jets \citep[e.g.][]{aloy99, saxton10}, and synthetic radio maps based on numerical radiative transfer calculations indicate that simple luminous knots are associated with the re-confinement, or pinching shocks \citep{saxton10}. Analysis of the re-confinement process provides an estimate of the shock separation in terms of the jet power and external pressure. Let $\Delta R$ be the separation between re-confinement shocks, $Q_{\rm jet}$ the jet power and $p_{\rm cocoon}$ the cocoon pressure. Analytic treatment of the re-confinement process, and numerical simulations of axisymmetric, hydrodynamic, relativistic jets indicate that, for a jet with Lorentz factor $\Gamma = 5$,
\begin{equation} \label{eqn:Delta_R}
\Delta R \approx 65 ~ \left( \frac{Q_{\rm jet}}{10^{46} ~ {\rm ergs/s}} \right)^{1/2} \left(  \frac{p_{\rm cocoon}}{10^{-11} ~ {\rm dyn/cm^2}} \right)^{-1/2} \> \mbox{kpc}~, 
\end{equation}
and is not sensitive to the jet Lorentz factor or jet opening angle \citep{komissarov98}. \citet{croston05} find that the typical magnetic field strength in the lobes of high power FRII radio galaxies is in the order of 10~$\mu$G, and that the particle energy density is typically a factor of a few greater than the magnetic energy density. This suggests that the typical lobe pressure is in the order of $10^{-11}$~dyn~cm$^{-2}$. Therefore, combining Equation \ref{eqn:Delta_R} with the de-projected knot separation ($50~\mbox{kpc} < D < 110~\mbox{kpc}$) and an assumed value for the cocoon pressure ($p_{\rm cocoon} \sim 10^{-11}$~dyn~cm$^{-2}$), we find that the re-confinement shock interpretation implies $Q_{\rm jet} \approx 10^{46}$~ergs~s$^{-1}$. 

Hydrodynamic simulations by \citet{saxton10} indicate that the ambient medium influences the distribution of jet knots resulting from reconfinement shocks. In a constant pressure atmosphere, the jet knots remain equally spaced along its length, but in an atmosphere with radially decreasing density, the knots become more closely spaced with distance along the jet. The nearly constant knot separation observed in PKS~0637--752 would therefore require the halo to have a core radius of hundreds of kpc, which is significantly larger than typically observed \citep{belsole07}. Furthermore, the re-confinement shock interpretation does not explain why the jet brightens in both the radio and X-ray bands at approximately 8 arcseconds from the core. If the difference in jet brightness between the inner and outer jet were due to a difference in jet power, or a difference in cocoon pressure, then we may expect to see this reflected as a difference in knot separation. Instead, we see that the distance between knots remains approximately constant, independent of the jet brightness. 
%However, an increase in flux density does not necessarily require a large increase in jet power. 

\subsubsection{Alternative Mechanisms}

A variety of mechanisms have been proposed to explain regular patterns in AGN jets \citep[e.\,g.][]{lapenta05, balsara92, bahcall95}. Most notably, Kelvin-Helmholtz (KH) instabilities \citep[e.\,g.][]{birkinshaw90} have been successfully employed to describe the emission patterns in pc-scale jets \citep[see review by][]{perucho12}, as well as the kpc-scale morphology of the jet in M87 \citep[e.\,g.][]{bicknell96, lobanov03}. A KH instability interpretation of the regular knot spacings in PKS~0637--752 would imply that the gross structure of the jet changes little over the region where this knot pattern is seen - in particular that the flow speed is constant - consistent with the IC/CMB model for the jet X-ray emission. We defer a more detailed investigation of the KH instability interpretation to an upcoming publication.

%We also mention here the soliton model of \citet{lapenta05} invoked to explain the regularly spaced knots in the kpc-scale jet of 3C303. In this model, the jet is electromagnetically dominated, and the knots represent a repeating pattern of ``magnetic bubbles" described by a soliton-like solution of the MHD equations. 

\subsection{Variation in the jet engine}  \label{sec:variation_in_the_jet_engine}

In the following sub-sections we consider models that invoke modulated activity of the jet engine to explain the semi-regular periodic appearance of the kpc-scale knots. To enable quantitative analysis, we must estimate the modulation timescale. Let us assume that the knots travel with Lorentz factor $\Gamma$, at an angle to the line of sight $\theta$. The corresponding Doppler factor is $\delta = [\Gamma (1 - \beta \cos \theta)]^{-1}$. Accounting for projection and light travel time effects, the distance between the jet knots ($D$) is related to the apparent separation between the knots ($D_{\rm app}$) via
\begin{equation}
D = \frac{D_{\rm app}}{\Gamma \delta \sin \theta} 
\end{equation}
If the knot velocity $\beta \approx 1$ then $\Gamma \delta \sin \theta \approx \beta_{\rm app}$ where $\beta_{\rm app}$ is the apparent transverse velocity of the kpc-scale knots. Hence,
\begin{equation} \label{eqn:D_app_limits}
\frac{D_{\rm app}}{\beta_{\rm app, max}} < D < D_{\rm app} \left[ \frac{1 - \beta \cos \theta}{\sin \theta}  \right]_{\rm max}
\end{equation}
The apparent transverse velocity of the pc-scale jet of PKS~0637--752 is $\beta_{\rm app} = 13.3 \pm 1.0$ \citep{edwards06, lovell00}
%, which implies a maximum viewing angle for the pc-scale jet of $\theta_{\rm max} < {\rm arcos} \left(  \frac{\beta_{\rm app}^2 -1}{\beta_{\rm app}^2 + 1} \right) \approx 9^{\circ}$
. We reasonably assume that the apparent transverse velocity of the kpc-knots is no greater than the apparent transverse velocity of the pc-scale knots, and therefore argue that $\beta_{\rm app, max} = 13.3$. The one-sidedness of the kpc-scale jet implies that the velocity remains at least mildly relativistic on kpc-scales. Based on the current radio and X-ray images, the jet to counter-jet brightness ratio is $\gtrsim 60$, implying that $\beta > 0.5$. As argued in Section \ref{sec:periodic_knots}, the jet viewing angle is $\theta > 4^\circ$. These limits, combined with equation \ref{eqn:D_app_limits}, imply that if the knots are associated with moving jet components, 
\begin{equation}
0.6~{\rm kpc} < D < 60~{\rm kpc}
\end{equation}
and given $1 < \beta < 0.5$, the inferred modulation period is
\begin{equation}
2 \times 10^3 ~ {\rm yr} < \tau < 3 \times 10^5 ~ {\rm yr}  \label{eqn:modulation_timescale}.
\end{equation}
%The lower end of this range is applicable if the jet remains highly relativistic on kpc-scales, as implied by the IC/CMB model of jet X-ray emission \citep{tavecchio00}.
The lower end of this range ($\tau \sim 2 \times 10^3$ yrs) is applicable if the jet remains highly relativistic on kpc-scales, as suggested by the IC/CMB model of jet X-ray emission \citep[][]{tavecchio00}. Given these constraints on the modulation timescale, we now consider two possible physical processes that could result in periodic modulation of the jet engine: accretion disk instabilities, and a binary supermassive black hole. \\

\vspace{-0.4cm}

\subsubsection{Accretion disk instabilities}

In the context of accretion disk instabilities, we emphasise a possible analogy between the jet of PKS~0637--752 and the quasi-periodic jet modulation in the microquasar GRS 1915+105 in the $\beta$ state \citep[][and references therein]{fender04}. The variations in jet output in GRS~1915+105, with a characteristic timescale of $\sim 30$ minutes, are believed to result from limit cycle behaviour in an unstable accretion disk \citep{janiuk02, janiuk11, tagger04}. Physical timescales are expected to scale linearly with the mass of the black hole, and therefore, the $\tau \sim 30$ minute oscillations in the jet of GRS~1915+105 \citep[$M_{\rm BH} \approx 14$~M$_\odot$;][]{greiner01} extrapolate to $\tau \sim 2 \times 10^3$~yrs for PKS~0637--752 \citep[M$_{\rm BH}\sim 5 \times 10^8$~M$_\odot$;][]{liu06}; consistent with the central engine modulation timescale inferred from the periodic structure in the jet.  

It has previously been suggested that thermal-viscous instability of the accretion disk may play a role in shaping the morphology of arcsecond-scale AGN jets \citep[e.\,g.][]{lin86, stawarz04a}. Two modes of thermal-viscous instability potentially operate in accretion disks around massive compact objects:  ionization instability and radiation pressure instability \citep[see e.\,g.][for a recent discussion]{janiuk11}. These instabilities are expected to result in quasi-periodic outbursts due to modulation of the accretion rate, $\dot{M}$. The ionization instability operates in the ``partially ionized zone" in the outer disk, where hydrogen transitions from being neutral to ionized. The characteristic ionization instability timescale for supermassive black holes with M$_{\rm BH} \approx 10^8$~M$_{\odot}$ is $\gtrsim 10^6$ yrs \citep{mineshige90, janiuk11}, which is longer than the outburst timescale required to explain the periodic structure in the jet of PKS~0637--752. 

%The ionization instability operates in the ``partially ionized zone" in the outer disk, where hydrogen transitions from being neutral to ionized. In this region, the opacity is strongly dependent on the temperature; the disk is unstable and transitions between a cold and hot state, resulting in modulation of the accretion rate. The ionization instability is widely accepted as the explanation for outbursts in dwarf novae and low-mass X-ray binary transients \citep{lasota01, cannizzo10, janiuk11}. \citet{lin86} showed that the ionization instability may also operate in the accretion disks around supermassive black holes in AGN. The characteristic instability timescale for supermassive black holes with M$_{\rm BH} \approx 10^8$~M$_{\odot}$ is $\gtrsim 10^6$ yrs \citep{mineshige90, janiuk11}, which is longer than the outburst timescale required to explain the periodic structure in the jet of PKS~0637--752. Furthermore, \citet{hameury09} argue that the ionization instability in AGN cannot produce large amplitude variations in the accretion rate, although this conclusion may be altered if the inner disk is truncated \citep{janiuk11}.  We therefore do not consider the ionization instability to be driving the quasi-periodic pattern in the jet of PKS~0637--752. 

The radiation pressure instability operates in the inner disk where radiation pressure dominates over gas pressure, and results in shorter timescale variability than the ionization instability \citep[see][]{janiuk11}. In the context of AGN, the radiation pressure instability has been proposed as an explanation for the intermittency of radio galaxies on short ($\sim 10^4$~yr) timescales, and in particular, the over-abundance of young radio sources in population studies \citep{reynolds97, marecki03, czerny09}. At high accretion rates (near Eddington), and black hole masses M$_{\rm BH} \sim 10^8 - 10^9$~M$_\odot$, the characteristic timescale between outbursts resulting from the radiation pressure instability is $\sim 10^3 - 10^4$~yrs \citep{merloni06, czerny09} --- consistent with the inferred modulation period in PKS~0637--752. %However, at high (near Eddingtion) accretion rates, the presence of a powerful jet may stabilise the disk against the radiation pressure instability \citep{janiuk11}. 

The interplay between the ionization and radiation pressure instability in AGN accretion disks may affect the disk behaviour \citep{siemiginowska96, janiuk11}. Variability over a range of timescales could potentially explain the variation in knot brightness between the inner and outer knots, as well as the variation in brightness between the knot and inter-knot regions. 

%The required modulation timescale 
%The regular periodic outbursts of the microquasar GRS 1915+105 have been interpreted in terms of the radiation pressure disk instability \citep[e.\,g.][]{janiuk02}.

An alternative instability, the magnetic flooding accretion-ejection instability, has been proposed to explain the $\beta$ state oscillations in GRS~1915+105, and may be applicable to AGN  \citep{tagger04}.

%\textit{See \url{http://iopscience.iop.org/0004-637X/737/2/69/pdf/apj\_737\_2\_69.pdf} The Physics of the ``heartbeat state" of GRS1915+105. In particular, follow-up the statement:
%``For example, it has been shown conclusively that 30 minute radio oscillations are Òbaby jetsÓ produced by ejection events in cycles like the $\beta$ state (Fender et al. 1997; Pooley \& Fender 1997; Eikenberry et al. 1998; Mirabel et al. 1998; Mirabel \& Rodrõguez 1999)."} 
%\textbf{But what is the $\beta$ state?}

%From Fender and Belloni 2004: ``With a general assumption that physical timescales scale with the mass of the black hole (e.g., Sams, Eckart \& Sunyaev 1996a and references therein), even one hour of disc-jet coupling in a 10 M$_\odot$ black hole probes an equivalent timescale of ?1000 years in a 10$^8$ M$_\odot$ SMBH."

%From Neilen, Remillard and Lee 2011: ``It is believed that many of these phenomenologically described variability classes, which are labeled with Greek letters (B00), are limit cycles of accretion and ejection in an unstable disk (Belloni et al. 1997a; Mirabel et al. 1998; Tagger et al. 2004; Fender \& Belloni 2004)."

%Finally, we note that such regularity in jet structure may not be compatible with an interpretation in terms of an instability, since such strong periodicity may not be expected in that case.  
 
\subsubsection{Binary supermassive black hole}\label{sec:BBH}
In this section we consider the possibility that the periodic structure in the jet of PKS~0637-752 arises from variations in the accretion rate driven by a binary black hole. 

Supermassive black hole (SMBH) binaries are expected to result directly from the merger of two massive galaxies, and may populate up to 10\% of local galaxies \citep{volonteri03,radiocensus}. However, they are notoriously difficult to detect; no definitive examples have been identified, although several potential binary precursor systems (i.\,e. dual active nuclei in a single galaxy at large separation) have been confirmed \citep{komossa03,rodriguez06,fabbiano11}. Periodicities in morphology, motion, and flux may arise from a bound supermassive black hole pair in a galaxy's centre \citep[e.\,g.][]{3c66b-2,macfadyen08}. Excess emission episodes correspond to a secondary SMBH's passage through disk-confined material (gas, stars) around the primary SMBH, inducing a heightened accretion rate at each traversal. Within this interpretation, the oscillatory signal indicates that the secondary SMBH's orbit is not in the same plane as the disk. Given a circular orbit, each observed knot may correspond to one disk passage event. However, if the orbit is elliptical, this model predicts that we are likely to observe double-peaked emission events corresponding to the initial and return traversal through the disk. The spacing of the sub-peaks will depend on the SMBH mass and ellipticity. OJ287, long theorised to be a SMBH binary, shows such structure in its light curve, emitting optical bursts at intervals of 11-12\,years, with two sub-peaks per emission cycle \citep{valtonen08}. Future high-resolution observations of PKS~0637--752 may reveal additional structure in the jet, such as double peaked knots.

Based on available information, we can estimate the properties of a putative binary system. The knot spacing corresponds either to $P$ (for an elliptical orbit) or $P/2$ (for a circular orbit), where $P$ is the orbital period of the binary system. This leads to periods of $2\times10^3<P<6\times10^5$\,years, and assuming a total system mass of $\sim 5 \times10^8{\rm M}_\odot$ (as estimated by \citealt{liu06} based on $H\beta$ luminosity and line-width), the corresponding orbital radius is $0.7\lesssim a \lesssim 30$~pc. This range in orbital radii implies that the largest possible angular separation of the binary lies in the range $\sim$0.1--5~mas. However, the current projected angular separation of the two SMBHs may be significantly smaller than the maximum angular separation, depending on the orbital orientation and phase.

The binary black hole interpretation cannot explain why the jet brightens in both the radio and X-ray bands at approximately 8 arcseconds from the core, and an additional mechanism, such as variation in the jet power, must be invoked to explain the sudden increase in jet brightness at 8 arcseconds. 

%If both supermassive black holes are radio-emitting, it is possible that two static (on timescales of decades) radio components would be visible through very long baseline interferometric imaging (VLBI). Space-based VLBI imaging and astrometric tracking of PKS~0637--752's core components has previously been performed \citep{lovell00, edwards06}. Inspection of the available images indicates the possible presence of a second stationary component within the 4 mas field in these VLBI experiments, which may be associated with a binary black hole companion. This cannot be definitively determined with the current data, but future high resolution observations at closely spaced intervals could reveal the presence of a stationary component associated with a companion SMBH.

\section{Discussion and Conclusions}
 
We have presented an 18~GHz ATCA map of the quasar PKS~0637--752, and identified a spectacular train of 9 quasi-periodic knots extending 11 arcseconds along the jet. 
We sought to address the question ``what physical process is responsible for the periodic knot structure?", and considered two classes of model: (A) those that involve a static pattern through which the jet plasma travels, and (B) those that involve quasi-periodic modulation of the jet engine. One model that falls under class A is the re-confinement shock interpretation. If the knots are associated with re-confinement shocks, the observed knot separation implies that the jet kinetic power is approximately 10$^{46}$ ergs~s$^{-1}$. However, the constant knot separation is not expected in a realistic external density profile. The re-confinement shock interpretation predicts a correlation between knot separation and jet kinetic power, which could be revealed in an imaging survey of a large sample of quasar jets showing regularly spaced knots. In this letter, we did not consider in detail alternative mechanisms under class A such as Kelvin-Helmholtz instabilities. We defer a more detailed investigation of the KH instability interpretation to an upcoming publication.

For models in Class B, the quasi-periodic structure in the jet is interpreted as arising from quasi-periodic modulation of the central engine. The inferred modulation period is $2 \times 10^3 ~ {\rm yr} < \tau < 3 \times 10^5 ~ {\rm yr}$. The lower end of this range is applicable if the jet remains highly relativistic on kpc-scales, as implied by the IC/CMB model of jet X-ray emission \citep{tavecchio00}.
%The wide range of periods is a consequence of the uncertainty in the Lorentz factor of the kpc-scale knots. 
This modulation timescale is consistent with the predicted radiation pressure instability timescale in the accretion disks of AGN. Indeed, we have drawn a direct comparison between the periodic structure in the jet of PKS~0637--752 and the quasi-periodic variation in jet activity seen in the microquasar GRS~1915+105, believed to result from limit cycle behaviour in an unstable accretion disk. Finally, variations in the accretion rate may be driven by a secondary black hole in orbit around a primary (i.\,e. more massive) black hole.  We estimated the orbital radius of a putative binary system to be $0.7\lesssim a \lesssim 30$~pc, which corresponds to a potentially-resolvable angular separation of $\sim$0.1--5~mas.  
 
%The discussion in this paper may be applicable to other quasar jets with quasi-periodic structure, such as those presented in \citet{bridle94}. 
 
\acknowledgments

The Australia Telescope Compact Array is part of the Australia Telescope which is funded by the Commonwealth of Australia for operation as a National Facility managed by CSIRO. 
DAS is supported by NASA contract NAS8-03060 and CXC grant GO9-0121B. A portion of research was carried out at the Jet Propulsion Laboratory, California Institute of Technology, under contract with the National Aeronautics and Space Administration.
  
{\it Facilities:} \facility{ATCA}

\end{document}